\documentclass[%
    preprint,
    superscriptaddress,
    amsmath,
    amssymb,
    aps,
    pre,
    table, 
    xcdraw
]{revtex4-1}

\usepackage{amssymb}
\usepackage{setspace}
\usepackage[utf8]{inputenc}
\usepackage{array}
\usepackage{natbib}
\usepackage{amsmath}
\usepackage{float}
\usepackage{multirow}
\usepackage{hyperref}
\usepackage{rotating}
\usepackage{adjustbox}
\usepackage{xcolor}
\usepackage[utf8]{inputenc}
\usepackage[T1]{fontenc}
\usepackage{textcomp}

\begin{document}

\title{Recovering link-weight structure in complex networks with weight-aware random walks}

\author{Adilson Vital Jr.}
\affiliation{Institute of Mathematics and Computer Sciences,
University of São Paulo, PO Box 369,
13560-970, S\~ao Carlos, SP, Brazil 
}

\author{Filipi N. Silva}
\affiliation{Observatory on Social Media, Indiana University, Bloomington, IN, USA
}

\author{Diego R. Amancio}
\affiliation{Institute of Mathematics and Computer Sciences,
University of São Paulo, PO Box 369,
13560-970, S\~ao Carlos, SP, Brazil 
}

\date{\today}

\begin{abstract}
Using edges weights is essential for modeling real-world systems where links possess relevant information, and preserving this information in low-dimensional representations is relevant for classification and prediction tasks. This paper systematically investigates how different random walk strategies -- traditional unweighted, strength-based, and fully weight-aware -- keeps edge weight information when generating node embeddings. Using network models, real-world graphs, and networks subjected to low-weight edge removal, we measured the correlation between original edge weights and the similarity of node pairs in the embedding space generated by random walk strategies. Our results consistently showed that weight-aware random walks significantly outperform other strategies, achieving correlations above 0.90 in network models. However, performance in real-world networks was more heterogeneous, influenced by factors like topology and weight distribution. Our analysis also revealed that removing weak edges via thresholding can initially improve correlation by reducing noise, but excessive pruning degrades representation quality. 
Our findings suggest that simply using a weight-aware random walk is generally the best approach for preserving node weight information in embeddings, but it is not a universal solution. 
\end{abstract}

\maketitle

\section{Introduction}
\label{chapter:introduction}

Weighted networks are widespread, appearing in domains including transportation, language, biology and finance~\cite{riascos2021random,amancio2016network}. In these networks, edge weights complement connectivity information by encoding the intensity, capacity, or strength of the connections between nodes. However, incorporating weights into network models introduces an additional layer of complexity. In some cases, edge weights are strongly correlated with the underlying topological structure -- for example, densely connected regions of the network may also feature stronger connections. In other cases, weights capture independent information that is not topologically encoded.

Graph embedding methods have gained significant traction as powerful tools for representing networks in low-dimensional vector spaces~\cite{dehghan2022evaluating}. Yet, despite their popularity, the role of edge weights in shaping the resulting embeddings remains underexplored. Given the additional complexity introduced by weights, a key question arises: to what extent can random walks and embedding techniques preserve weight-related information, and under what conditions is this information recoverable?

Some studies have shown that incorporating edge weights can enhance performance in downstream tasks, as seen in models such as ARGEW~\cite{kim2023node}, Node2Vec+\cite{liu2023accurately}, and ProbWalk~\cite{wu2021probwalk}. These models adapt random walks to account for edge weights, demonstrating improvements in node representation learning. However, many other approaches still rely on unweighted, binary representations of the network \cite{liu2023accurately}, often ignoring the rich structural and organizational information embedded in the weights~\cite{fardet2021weighted}.
Research focusing on random walks -- particularly in unweighted settings -- has highlighted their ability to reconstruct network topology and recover structural metrics~\cite{vital2024comparing, guerreiro2024identifying, guerreiro2024recovering}.

In this work, we build upon different edge-weighted biased walks \cite{wu2021probwalk}. In particular we compare the strategies: a walk biased by strength of the destination node and a standard uniform random walk. We encode the nodes into an embedding representation using skip-gram and negative sampling by following the weighted adaptation of Node2Vec \cite{grover2016node2vec} and Node2Vec+ \cite{liu2023accurately}.
Our primary goal is to investigate how well weighted random walks and node embeddings can retain information about the original edge weights, and which factors influence this relationship. These insights could inform future applications of graph learning where edge weights are a critical component during training.

To this end, we conducted three sub-experiments: (i) a sensitivity analysis using synthetic networks by varying parameters that control network generation and walk configuration, evaluating the correlation between cosine similarity in the embedding space and the original weights; (ii) an application-focused test using real-world networks and varying walk parameters; and (iii) a robustness analysis in which complete networks are pruned by iteratively removing the lowest-weight edges, examining how correlation behaves under progressive sparsification.

Our findings indicate that, in some cases, even unweighted walks can recover weight-related information, particularly in networks where edge weights are topologically aligned, such as geographic graphs. Nonetheless, weighted walks consistently yielded stronger correlations across both synthetic and real-world datasets. Performance drops were observed in networks with highly skewed weight distributions, where a small number of nodes concentrate most of the edge weight. 
In real-world networks, where structure is more complex, correlation is harder to achieve, especially in power-law graphs. Finally, the pruning experiment revealed the existence of an optimal threshold of edge removal that initially improves correlation before leading to model degradation.

\section{Related Works}
\label{chapter:related}

Embeddings are widely used in textual and graph domains as a powerful tool for transforming discrete and unstructured data -- such as words or nodes -- into dense vectors that encode structural or semantic relationships~\cite{vital2024predicting,vital2024comparing}.
%

A pivotal advancement in word embedding was the introduction of Word2Vec~\cite{mikolov2013efficient, mikolov2013distributed}, which represented a significant improvement over earlier methods such as n-gram models~\cite{bengio2003neural, schwenk2007continuous}. Although effective, these earlier approaches faced substantial scalability challenges. 
In particular, n-gram models struggled to handle vocabularies with hundreds of millions of words within reasonable training times and failed to produce high-quality embeddings with more than 100 dimensions.
This limitation led to the development of distributed representations, where each word is represented as a dense vector in a continuous space.
These embeddings capture both local and global contextual information, encoding semantic and syntactic properties that can be effectively leveraged in downstream tasks such as sentiment analysis, text classification, and regression~\cite{ma2015using, al2019use, jatnika2019word2vec, kurnia2021classification}.

Word2Vec overcame these limitations by introducing a new neural training paradigm. Each unique word in the vocabulary is first converted into a one-hot encoded vector with dimensionality equal to the vocabulary size. During training, for a given sentence, a word is selected as the center word, and a fixed-size window of surrounding words is defined as its context. 
Depending on the architecture -- continuous Bag-of-Words (CBOW) or Skip-Gram -- the model either predicts the center word from its context or predicts the context words from the center word, respectively. 
The input one-hot vectors are passed through a single hidden layer, and the resulting activations serve as word embeddings. The weight matrix between the input and hidden layer effectively projects the high-dimensional one-hot vectors into a lower-dimensional embedding space, with the dimensionality of the hidden layer determining the size of the resulting word vectors.

In Word2Vec, the training process is optimized by discarding extremely common words (stop words) and filtering out rare ones. This allows the model to focus on learning meaningful contextual relationships between words, thereby enhancing the quality of the embeddings~\cite{goldberg2014word2vec}. To accelerate training, hierarchical softmax \cite{morin2005hierarchical, mikolov2013efficient} is employed, replacing comparisons over the full vocabulary with comparisons against the log of the vocabulary size. 
A later improvement, inspired by Noise Contrastive Estimation \cite{gutmann2012noise, mnih2012fast}, introduced negative sampling, which compares a given word to a small, fixed number of randomly sampled noise words (typically fewer than 20), even when training on large datasets.

Drawing inspiration from the success of Word2Vec, DeepWalk~\cite{perozzi2014deepwalk} emerged as a graph embedding framework. It tackles analogous limitations in graph representation learning, paralleling Word2Vec's impact on natural language processing. It outperformed traditional graph embedding approaches such as multidimensional scaling~\cite{cox2000multidimensional}, IsoMap \cite{tenenbaum2000global}, and Laplacian Eigenmaps \cite{belkin2001laplacian}, particularly in scalability. 
DeepWalk can be applied to large graphs and produces low-dimensional, continuous, task-agnostic vectors that preserve structural and social relationships among nodes. The method involves applying truncated random walks starting from each node, where the probability of visiting a neighboring node is uniformly distributed. The collected sequences are then used to train a Skip-Gram model with hierarchical softmax.

DeepWalk pioneered the use of unbiased random walks to gather node sequences, much like word sequences in natural language processing (NLP). These sequences are then fed into neural models to generate embeddings.
Subsequently, GraRep~\cite{cao2015grarep} and LINE~\cite{tang2015line} were proposed. With distinct approaches, GraRep leveraged matrix factorization to capture global structural information, while LINE applied an objective function aimed at preserving both first and second-order node proximities. Both models yielded impressive results, exceeding those of DeepWalk. 

Walk-based graph embedding models regained prominence with the introduction of Node2Vec~\cite{grover2016node2vec}, which used biased random walks controlled by parameters that simulate breadth-first (BFS) or depth-first (DFS) sampling strategies. This flexibility enabled the model to learn both homophilic (social) and structural (role-based) similarities among nodes, leading to strong performance in tasks such as link prediction \cite{de2018combining}, node classification \cite{grover2016node2vec}, and regression \cite{makarov2019link}.

The introduction of biased random walks that capture structural information enabled the exploration of random walks in weighted networks. Reweighting techniques \cite{kosmatopoulos2021random} were proposed to assign new weights to edges based on node similarity~\cite{vital2022comparative, hasan2011survey}.
%
Data augmentation models like ARGEW~\cite{kim2023node} inflate the number of walk sequences after generation, using the maximum edge weight for each node to produce robust and well-performing models. Two additional models, Node2Vec+~\cite{liu2023accurately} and ProbWalk~\cite{wu2021probwalk}, directly incorporate edge weights into transition probabilities. Node2Vec+ extends the original Node2Vec by integrating weights while retaining the local-global walk behavior. ProbWalk, on the other hand, uses edge weights directly in the transition probabilities, giving higher-probability transitions to higher-weight edges.

Node2Vec+ and ProbWalk generate walk sequences based on edge-weighted transitions, which are then fed into Skip-Gram models with negative sampling for embedding generation. These models demonstrate high stability under varying hyperparameter configurations and outperform predecessors such as DeepWalk, LINE, and Node2Vec in multiple downstream tasks. However, it is important to note that the primary focus of these models is on their application and performance in downstream tasks, rather than on analyzing the effectiveness of random walks in extracting structural information from networks. Other studies have narrowed their focus to comparing biased walks and their ability to capture topological information through the generated sequences \cite{guerreiro2024identifying, guerreiro2024recovering}. \cite{vital2024comparing} analyzed different biased walks in downstream tasks and showed that even walks with opposite exploration strategies are highly correlated and yield similar performance.

\section{Objectives and Goals}
\label{chapter:goal}

In many real-world networks, edges are associated with weights that convey domain-specific semantic meanings, like distance in geographic networks or proximity in social networks. In addition to their semantic value, these weights encapsulate crucial structural information about network organization and dynamics \cite{fardet2021weighted}.
While prior work has progressively improved weighted models, a fundamental understanding of their mechanics is lacking. There has been no systematic comparison of how different edge-weighted random walk approaches contribute to the node embedding process or their efficacy in translating edge weight information into the resulting vector space. The research focus has largely been on evaluating utility in downstream tasks like link prediction, neglecting a deeper investigation into how, precisely, weight information is encoded. This study addresses this oversight by providing a comprehensive analysis of different random walk strategies across a diverse set of networks.

The primary objective of this study is to analyze random walks on weighted graphs to assess whether node sequence–based embeddings are sensitive to edge weights and walk dynamics. 
Specifically, we evaluate whether embedding models that incorporate edge-weight–guided random walks can effectively \emph{encode weight information into the resulting node embedding representations}, thereby demonstrating awareness of the graph weighting information. 
We further investigate how the encoding ability is influenced by several factors: the underlying (unweighted) network structure, the type and bias of the random walk employed, and intrinsic characteristics of the graph such as directionality, sparsity, or density. Additionally, we examine how the presence of missing or filtered data -- common in real-world network analysis -- affects the embeddings’ capacity to preserve meaningful weight information.

By addressing these goals, this study aims to advance our understanding of the interplay between random walk dynamics and weighted graph structures. We seek to provide insights into how effectively walk-based embeddings can capture weight information, thus informing broader applications in network analysis and representation learning.

\section{Methodology}
\label{chapter:methodology}

Figure \ref{fig:methodology} illustrates the methodology employed in this paper. Each step is summarized below: 
\begin{figure}
\centering
\includegraphics[scale=0.75]{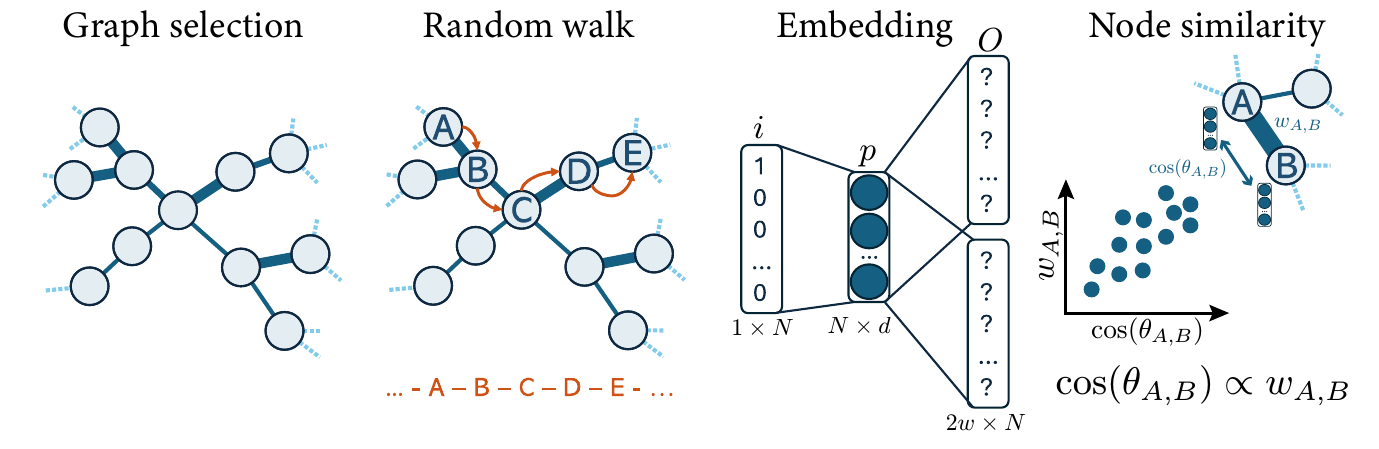}
\caption{Example of the methodology employed to analyze embeddings generated via biased random walks in weighted networks. The process begins with graph selection, followed by the execution of random walks to yield sequences of nodes. These sequences subsequently serve as input for generating node embeddings. The resulting embeddings are then subjected to analysis through the proposed experiments, for instance, by leveraging them to measure node similarity. }
\label{fig:methodology}
\end{figure}

\begin{enumerate}

\item \emph{Graph Selection}: our study on biased random walks in networks is conducted on a diverse set of networks with varying structures, sources, and weight distributions. These networks primarily come from three categories: synthetic graph models, real-world networks, and complete graphs.

For graph models, we opted for those that different represent common characteristics of the real-world networks. This includes models that preserve community structure, locality and scale-free degree distributions. This allows us to control the characteristics of the networks and evaluate how structural and weight-related properties influence the embedding process. For real-world networks, we opted for a diverse set of weighted networks -- from different contexts and with weights indicating different properties. Finally, we generate a set of weighted complete graphs where edges are progressively removed based on their lower weight values. This step helps us evaluate how well the embeddings generated via biased random walks adapts to changes in \emph{connectivity} and \emph{weight} distribution.

\item \emph{Random Walk}: we employed three distinct random walk strategies on each graph. Two of these strategies incorporated edge weights as a bias for transition probabilities, thereby increasing the likelihood of traversing higher-weight connections. The third strategy, a pure random walk, disregarded edge weights entirely, functioning as a control to evaluate the influence of weight information during the embedding process.

\item \emph{Embedding}:  by performing several realizations of the random walks, we obtain a collection of sequences of nodes. Then we apply the same logic behind Node2Vec, in which such sequences are fed into the Word2Vec, traditionally used in language processing. To convert the node sequences into embedding vectors, we employ both Skip-Gram and Continuous Bag-of-Words (CBOW) architectures. Given that nodes within a graph function analogously to words in a sentence, the resulting embeddings effectively capture not only individual node properties but also their neighborhood relationships and the graph structure. 

\item \emph{Node Similarity}: the primary objective of this study is to evaluate whether random walks can effectively encode edge weight information into node embeddings. To assess this, we compute the cosine similarity between all connected node pairs in the embeddings generated for each combination of graph and random walk technique. We then measure the Pearson correlation between the original edge weights and their corresponding cosine similarities in the embedding space. A high absolute correlation suggests that the embedding process successfully preserves weight information.

\end{enumerate}

\subsection{Graph Selection}

\subsubsection{Weighted Graph Models}

We used four well-known graph models, modified to include weighted edges: the Erdős–Rényi model, the stochastic block model (SBM), the Waxman model, and the Barabási–Albert model. Each graph varied in size, following powers of 2, ranging from 128 to 4096 nodes, and the average node degree between 2 and 32, using the same logic. To ensure statistical robustness, 10 random instances of each graph configuration were generated. This approach allows us to assess the consistency of our findings across multiple graph models and configurations. 
Figure \ref{fig:modelgraphs} illustrates the topology of the main network models used. Dark blue indicates edges with higher weights and light blue represents those with lower weights. The models are described below: 

\begin{figure}
\centering
\includegraphics[scale=0.50]{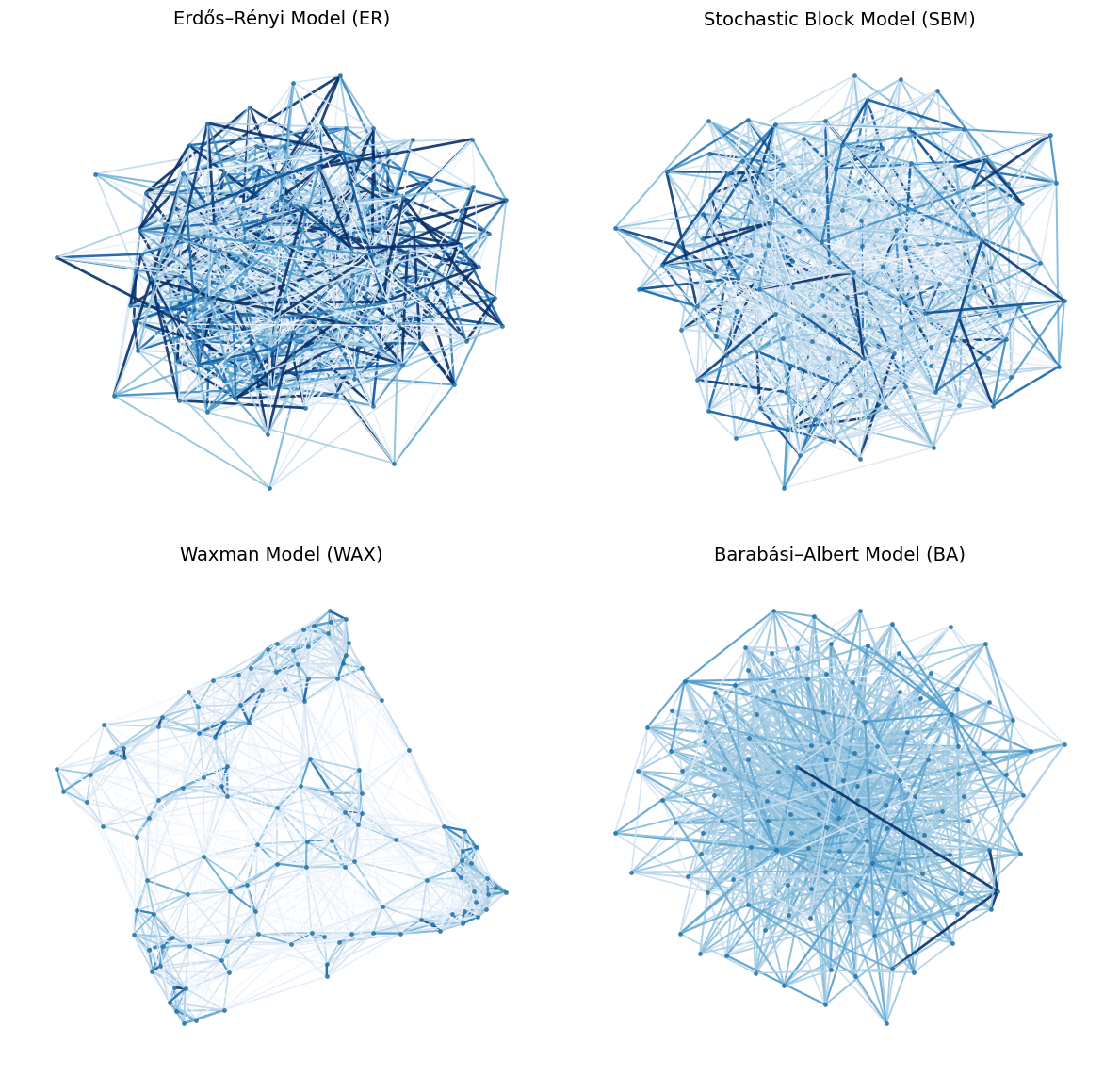}
\caption{Visualization of the four different graph models with weighted edges, each containing 128 nodes and an average degree of 16. The Erdős–Rényi (ER) model assigns edge weights following an exponential distribution, while the Stochastic Block Model (SBM) determines weights based on the probability of connections within and between node communities. The Waxman (WAX) model incorporates spatial constraints, with edge weights influenced by the geometric distances between nodes. Finally, the Barabási–Albert (BA) model represents a weighted scale-free network, where edge weights are proportional to the degrees of the connected nodes.}
\label{fig:modelgraphs}
\end{figure}

\begin{itemize}

    \item \emph{Erdős–Rényi Model (ER):} The Erdős–Rényi model~\cite{erdos59a}, also known as the random graph model, is defined by two parameters: $n$ and $p$. The model initializes a graph with $n$ nodes, and each pair of nodes is connected with probability $p$. 
    We assigned edge weights randomly based on three different functions: a uniform distribution, with values ranging from 0.1 to 1; a normal distribution, with a mean of 0.55 and a standard deviation of 0.15; and an exponential distribution with a scale parameter of 0.45. 
    
    \item \emph{Stochastic Block Model (SBM):} the Stochastic Block Model~\cite{HOLLAND1983109} extends the ER model by introducing community structure. In this model, $P_{in}$ represents the probability of an edge existing between two nodes within the same community, and $P_{out}$ represents the probability of an edge existing between nodes in different communities. We fixed the number of communities to 5 and tested various combinations of $P_{in}$ and $P_{out}$ to achieve the desired average node degree for a given graph size. We always ensured that $P_{in} > P_{out}$, reflecting a higher probability of connections within the same community than between different communities. Edge weights were assigned as random probabilities corresponding to the likelihood of connection between two nodes.

    \item \emph{Waxman Model (WAX):} The Waxman model~\cite{12889} generates geometric graphs where random distances are assigned to all pairs of nodes. Using the parameters $\alpha$ and $\beta$, the probability of an edge existing between two nodes is calculated as:  
    \begin{equation} 
    \mathbb{P}(e_{uv} = 1) = \alpha \exp\left(-\frac{d(u, v)}{\beta}\right).
    \label{eq:waxprob}
    \end{equation}
    Here, $d(u, v)$ represents the distance between nodes $u$ and $v$. The probability decreases as the distance increases, making closer nodes more likely to connect. The calculated probabilities were then assigned as weights to the existing edges.

    \item \emph{Barabási–Albert Model (BA):} The Barabási–Albert Model~\cite{Baraba_si_1999} incrementally grows a graph by adding nodes until it reaches the maximum size of $n$ nodes. At each step, a new node is connected to $m$ existing nodes, where nodes with more connections have a higher probability of being selected. This process produces networks with a power-law degree distribution, where a small number of nodes have many connections, while most nodes have few.  
    For our experiments, we used two versions inspired by~\citep{PhysRevLett.86.5835}: the Weighted Scale-Free (WSF) and the Weighted Exponential (WE).
    In the WSF version, we began with a regular scale-free graph. Weights were then assigned proportionally to the degree of the originating node, with the constraint that the weights for each node summed to 1.
    For the WE version, weights were updated dynamically during each iteration. When a new node was added, it connected to $m$ existing nodes. The weights were then recalculated proportionally to the degree of each chosen node after these new connections were made. This approach ensures that nodes that already have a higher degree at the time of connection receive stronger new connections.

\end{itemize}

\subsubsection{Real Graphs}

For the real-world networks, we used the Netzschleuder network catalog~\cite{tiago_p_peixoto_2020_7839981}, selecting 11 weighted graphs that are bidirectional -- meaning the origin and destination nodes are interchangeable -- and unipartite, indicating that all nodes are of the same type. We also restricted the selection to networks with fewer than 5,000 nodes. Additionally, we chose graphs with diverse structural properties, such as varying numbers of nodes, edges, and average node degrees, as well as a mix of network domains, including social, biological, and textual datasets. This diversity ensured a practical and representative testbed for evaluating our method under realistic conditions, while capturing a broad range of graph characteristics. Additional details can be found in Table \ref{tab:realnetworksfeatures}.

\begin{table}[H]
    \centering
    \caption{Real-world weighted networks used in this study. Each network is characterized by the number of nodes ($N$), number of edges ($E$), and the average node degree ($\langle k \rangle$).}
    \label{tab:realnetworksfeatures}
\begin{tabular}{|l|c|c|c|}
\hline
\textbf{Graph Name}                   & $N$ & $E$ & $\langle k \rangle$ \\ \hline
{Sp High School Diaries}       & 120            & 348            & 5.8                       \\ \hline
{Celegansneural}               & 297            & 2148           & 14.5                      \\ \hline
{Fao Trade}                    & 214            & 9441           & 88.2                      \\ \hline
{Celegans 2019 Hermaphrodite}  & 446            & 4210           & 18.9                      \\ \hline
{Celegans 2019 Male}           & 559            & 4560           & 16.3                      \\ \hline
{Faculty Hiring US Academia}   & 3284           & 52163          & 31.8                      \\ \hline
{Cintestinalis}                & 205            & 2624           & 25.6                      \\ \hline
{Budapest Connectome All 200k} & 1015           & 105293         & 207.5                     \\ \hline
{New Zealand Collab}           & 1463           & 4246           & 5.8                       \\ \hline
{Bible Nouns}                  & 1707           & 9059           & 10.6                      \\ \hline
{Netscience}                   & 379            & 914            & 4.8                       \\ \hline
\end{tabular}
\end{table}

\subsubsection{Complete Graphs} \label{sec:complete}

To evaluate how the removal of low-weight edges impacts the ability of random walks to capture weight information, we conducted a controlled experiment using completed graphs -- i.e., fully connected networks where every node is linked to every other node.

We began by generating a random feature matrix, where each row represents a node and each column corresponds to a feature dimension. The number of rows is equal to the graph size (i.e., the number of nodes), and the number of columns defines the feature dimensionality of each node. Using this matrix, we calculated the cosine similarity between every pair of nodes to assign weights to the edges. This resulted in a fully connected weighted graph, where the edge weight between two nodes reflects the similarity of their feature vectors.

Next, we applied a percentile-based thresholding procedure to progressively remove edges with low similarity scores. For a given threshold (e.g., 0.3), all edges with weights below the 30th percentile were removed from the graph. This allowed us to analyze how reducing noise -- by removing weak or less meaningful connections -- affects the quality of node embeddings.

\subsection{Random walks}

To prepare input for the embedding model, node sequences were generated by performing random walks on the graph. A random walk is a dynamic process where an initial node is selected as the starting point, and subsequent nodes are chosen iteratively from its neighbors based on a predefined selection function. This process continues until the maximum walk length is reached, completing the node sequence. 

To comprehensively explore the network structure, multiple random walks are initiated from each node in the graph, ensuring diverse and representative coverage of the network. The choice of the selection function plays a critical role in shaping the walks. While a fully random selection function treats all neighbors equally, alternative approaches introduce biases based on node properties such as degree. For example, a degree-biased random walk preferentially selects hubs and high-degree nodes, while an inverse-degree-biased walk favors isolated or low-degree nodes. 

In this work, three types of random walks were employed to enable a comparative analysis of their effects on downstream tasks: the traditional random walk, the strength-based walk and the weighted random walk:

\begin{itemize}

    \item \emph{Traditional Random Walk:} a well-known method where the probability of transitioning to a neighboring node is random, being dependent only on the number of connections that the origin node has. In other words, the probability of transitioning from node $u$ to node $v$ is $P(u \to v) = k_u^{-1}$, where $k_u$ is degree of $u$.
    %
    %
    %
    This random walk is purely random and does not depend on the edge weight, node strength, or the degree of the target node ($v$). However, it serves as a good benchmark for understanding what a non-weighted walk can embed from the network's structure. 
    
    \item \emph{Strength-Based Random Walk:} the probability of transitioning to a neighboring node is proportional to its strength (i.e., the sum of weights of edges connected to the node). Nodes with higher strength have a higher likelihood of being visited. The transition probability is given by: 
    \begin{equation} 
        P(u \to v) = \frac{s_v}{\sum_{w \in \mathcal{N}(u)} s_w},
    \label{eq:SRW}
    \end{equation}
    where $s$ is the strength of the node and $\mathcal{N}(u)$ is the set comprising the neighbors of node $u$. 
    The weight of edge linking $u$ and $v$ is relevant to the probability. However, this weight does not play a significant role on its own because the neighbors may have other strong connections that increase their chance of being transitioned to.
    
    \item \emph{Weighted Random Walk:}  the transition probability is directly proportional to the edge weight. This ensures that edges with higher weights are more likely to be traversed, emphasizing the importance of weights in the graph structure. The transition probability is given by
    \begin{equation} 
        P(u \to v) = \frac{w_{uv}}{\sum_{w \in \mathcal{N}(u)} w_{uw}},
    \label{eq:WRW}
    \end{equation}
    where $w_{uv}$ is the  weight associated with the edge linking nodes $u$ and $v$.  
    
\end{itemize}

\begin{figure}
\centering
\includegraphics[scale=0.65]{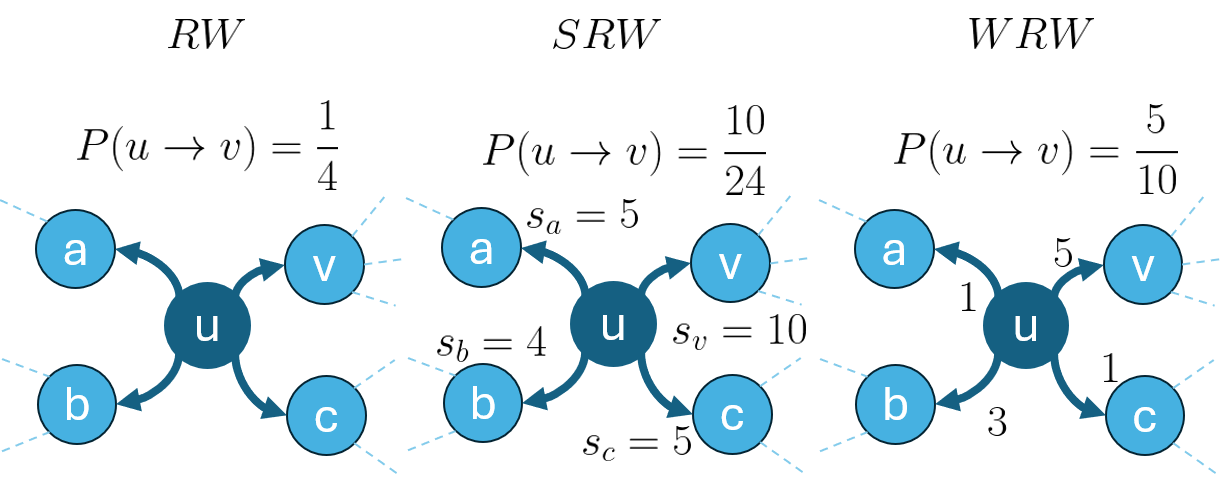}
\caption{Representation of the three different biased random walk models. The first model, Random Walk (RW), uses the degree of the initial node as the transition probability bias, leading to an equal probability of selecting any neighbor. The second model, Strength-based Random Walk (SRW), biases transitions based on the strength of the destination node, where strength is defined as the sum of the weights of all edges connected to a node. The third model, Weighted Random Walk (WRW), directly incorporates edge weights as biases, making transitions more likely along higher-weight edges.}
\label{fig:randomwalks}
\end{figure}

Examples of how to calculate the random walks are shown in Figure \ref{fig:randomwalks}. 

To evaluate how different network and walk parameters influence the correlation between cosine similarity and edge weights, we systematically repeated the experiment across multiple configurations, varying one parameter at a time while keeping the others fixed. For graph size, we varied the number of nodes from 16 to 1024, while maintaining 16 features, 16 walks per node, and a fixed walk length of 128. For the number of features, we tested values ranging from 2 to 32, with the graph size fixed at 512, 16 walks per node, and a walk length of 128. For walks per node, we varied the value from 2 to 32, using a graph size of 512, 16 features, and a walk length of 128. Finally, for walk length, we increased it from 16 to 256, while keeping the graph size at 512, the number of features at 16, and the walks per node at 16.

\subsection{Embedding Creation}

Similar to how sentences are created in natural language processing, random walks generate sequences of nodes. We used the skip-gram model -- a popular word embedding technique -- to convert these sequences into a numerical representation for embedding creation.
The skip-gram model learns vector representations (embeddings) for nodes by predicting the context of each node within a sequence. Basically each node in the graph is firstly represented by a one-hot vector with size $N$, being $N$ the graph size, where all elements will be zero except the nth element with value 1 representing the $n$-th node in the graph. Then, a moving window of length $w$ will go through the node sequence, with the central one being used as input for a neural networks model with $\kappa$ neurons in the hidden layer to predict the nodes surrounding it, before and after. This process will repeat until go through all the sequence, for all node sequences. After convergence of the training, the weights that the $\kappa$ neurons learned are used to project each one hot vector representing a node into an embedding vector of size $\kappa$. 

In this context, the model captures the structural and relational information of each node within the graph. The resulting embeddings encode both local (neighborhood-level) and global (graph-level) structural features. These embeddings serve as a powerful tool for analyzing the graph's characteristics and comparing nodes based on their learned representations. 

\subsection{Weight Correlation Analysis}

To test the hypothesis that embedding similarity reflects the weights of the original network, we computed the cosine similarity between the node embeddings. Our goal was to evaluate how effectively the random walk process encodes edge weights into these embeddings. By comparing the cosine similarities to the original edge weights using the Pearson correlation coefficient, we assessed the extent to which the embeddings preserve the graph’s weighted structure.

\section{Results and Discussion}
\label{chapter:results}

\subsection{Parameter Sensivity Analysis}

In the first experiment, we conduct a parameter-sensitive analysis by applying weighted random walks to weighted variants of well-established graph models. 
As in related studies \cite{barbour2024evaluating, fardet2021weighted}, we vary the parameters of the graph construction and walk configuration \cite{liu2023accurately} to observe how specific structural or stochastic features influence the resulting correlation between edges weight and embeddings similarity. This setup allows us to analyze weight encoding patterns under controlled, synthetic conditions -- providing valuable insight that is often obscured in the complexity of real networks.

\subsubsection{Graph Size}

\begin{figure}
\centering
\includegraphics[scale=0.55]{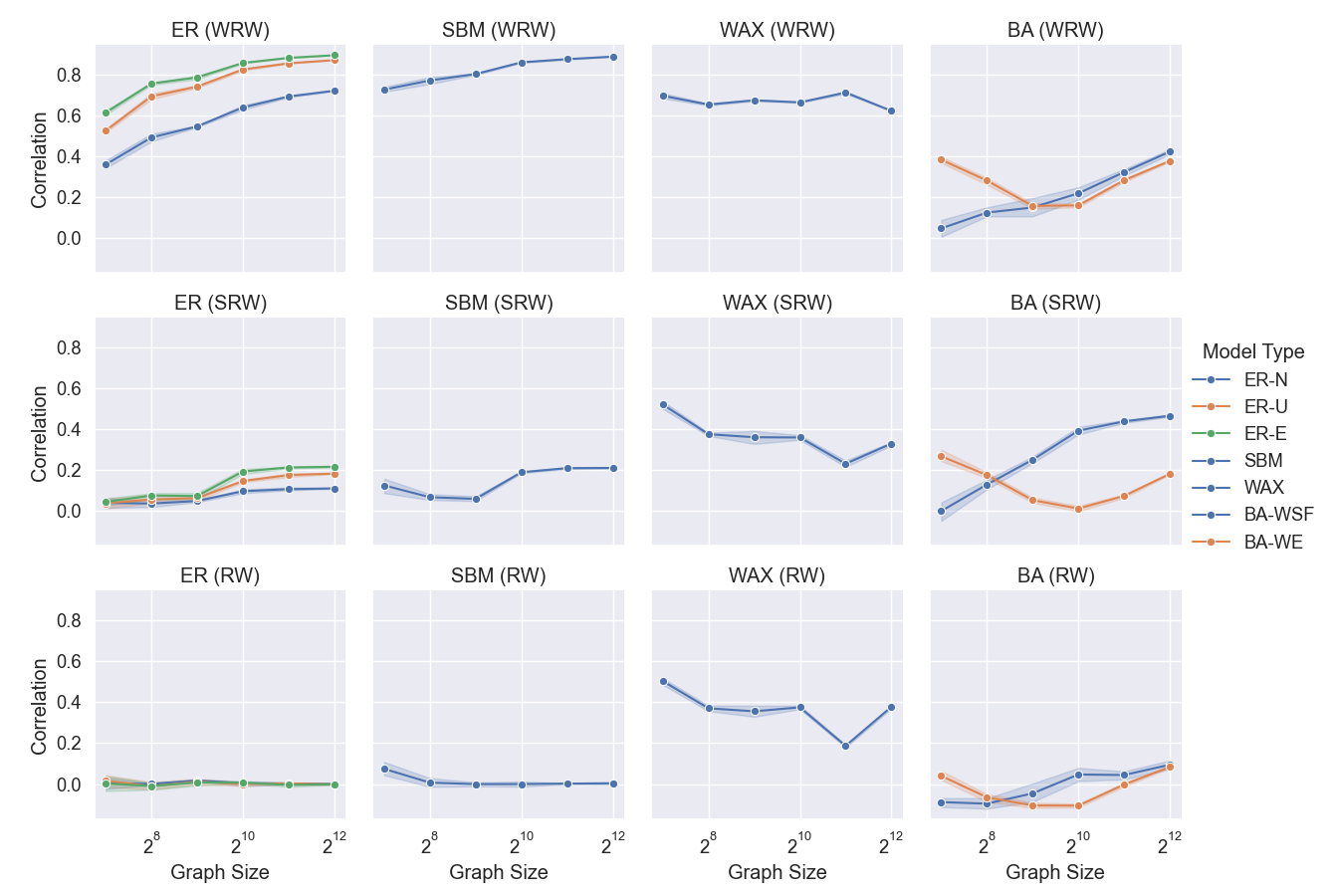}
\caption{Pearson correlation between cosine similarity of node embeddings and original edge weights across different graph models and random walk strategies. This analysis evaluates how well three types of random walks (WRW, SRW, RW) capture edge weight information in four graph models (ER, SBM, WAX, BA). Correlations were computed across multiple network instances, analyzing trends as graph size increased from 128 to 4096 nodes, with fixed parameters: walk length of 128, 16 walks per node, and an average node degree of 16.}
\label{fig:model_graph_size}
\end{figure}

The first parameter analyzed in our study is the graph size. 
Figure \ref{fig:model_graph_size} presents the results of fixing the average node degree to 16, and doing 16 walks per node of length 128, while increasing the graph size in powers of two, ranging from 128 to 4096 nodes, with the rows representing the walk types, and the columns the graph models. 

The first row in Figure \ref{fig:model_graph_size}, corresponding to the traditional random walk (RW), shows a very low correlation for the ER, SBM and BA graphs. For the first two, this behavior is expected, as these models rely on random probabilistic connections without an inherent structure that embeds weight relationships. However, a different pattern is observed in the BA and WAX models, where the graph structure influences weight assignment. In the BA model, higher weights are assigned to well-connected nodes, reinforcing its scale-free nature, with the structure embedding this information which can be observed that the plot initially exhibits a slightly negative correlation that increases as the graph size grows. In WAX, weight assignment is based on distance, meaning shorter distances increase the likelihood of edge formation, while very long distances remain rare, effectively simulating transportation or city connectivity networks. As a result, even though the traditional random walk does not incorporate information about edge weights, it still achieves a correlation of approximately 0.40, outperforming the other models. This indicates that WAX is able to effectively embed weight information into the network structure.

The strength-based random walk (SRW) leads to an overall increase in correlation compared to RW. ER and SBM had a similar behavior with correlation remaining quite low until reaching 512 nodes, after which a slight increase stabilizes around 0.2 for larger graph sizes. WAX maintains a consistent behavior across different graph sizes, with correlation values ranging between 0.2 and 0.6. However, the most significant change occurs in BA, where the two variations exhibit distinct behaviors. The weighted exponential model (BA-WE) follows a U-shaped curve, where correlation initially decreases as graph size increases up to 1024 nodes, followed by a subsequent rise. In contrast, the weighted scale-free model (BA-WSF) displays a steady increase, starting from an almost zero correlation at 128 nodes and reaching approximately 0.5 at 4096 nodes. This difference can be explained by the way weights are assigned in these models. In BA-WSF, weights are assigned proportionally to node degree, meaning high-degree nodes naturally accumulate larger weights. Since node strength is the sum of edge weights, high-degree nodes also exhibit high node strength, increasing their likelihood of being visited during the walk. As a result, these nodes appear more frequently in the sampled sequences, reinforcing their importance in the embedding space and leading to higher cosine similarities between edges connecting them. In contrast, BA-WE dynamically assigns weights in proportion to the degree during each iteration, which means that nodes with the highest degrees may shift throughout the network’s growth. This dynamic reassignment weakens the original weight correlation, leading to a temporary drop before correlation recovers at larger sizes.

The weighted random walk (WRW) achieves the best results across all models except for BA, demonstrating a strong correlation between learned node embeddings and the original edge weights. In ER, all three weight distributions -- uniform (U), exponential (E), and normal (N) -- perform well, with exponential and uniform distributions achieving correlations above 0.9. Similarly, SBM follows a trend comparable to ER, with correlation values ranging between 0.7 and 0.9. The WAX model continues to exhibit strong and stable performance, with correlation values between 0.6 and 0.7. Notably, the drop observed at size 2048 in the other walks instead manifests as a peak in WRW, yielding the highest correlation among all graph sizes. BA exhibits a pattern similar to SRW, but with a key distinction: both sub-models exhibits a similar uptrend. 
Another issue is that increasing the number of nodes can lead to a greater number of hubs. These hubs introduce noise to the random walk, which skews the distribution of visited nodes and, consequently, results in poorly represented embeddings.

Overall, the traditional random walk (RW) demonstrates poor performance in terms of correlation, which aligns with expectations since it does not incorporate edge weights in its sampling process. The only exception to this trend is the WAX model, where the underlying structure inherently captures distance relationships, resulting in a consistently positive correlation. The incorporation of node strength in SRW slightly improves correlation, particularly in BA-WSF, where a steady increase is observed as graph size grows. However, WRW outperforms all other approaches, achieving correlations exceeding 0.9 in some cases, demonstrating its superior capability in encoding edge weights. Additionally, BA achieves its highest correlation with SRW, which is consistent with the way its weights are embedded within the network’s scale-free structure. These findings highlight the effectiveness of weight-aware random walk strategies in preserving edge weight relationships within node embeddings, making them particularly useful for applications where both graph structure and edge weights carry meaningful information, such as transportation modeling, recommendation systems, and financial networks.

\subsubsection{Average Node Degree}

\begin{figure}
\centering
\includegraphics[scale=0.5]{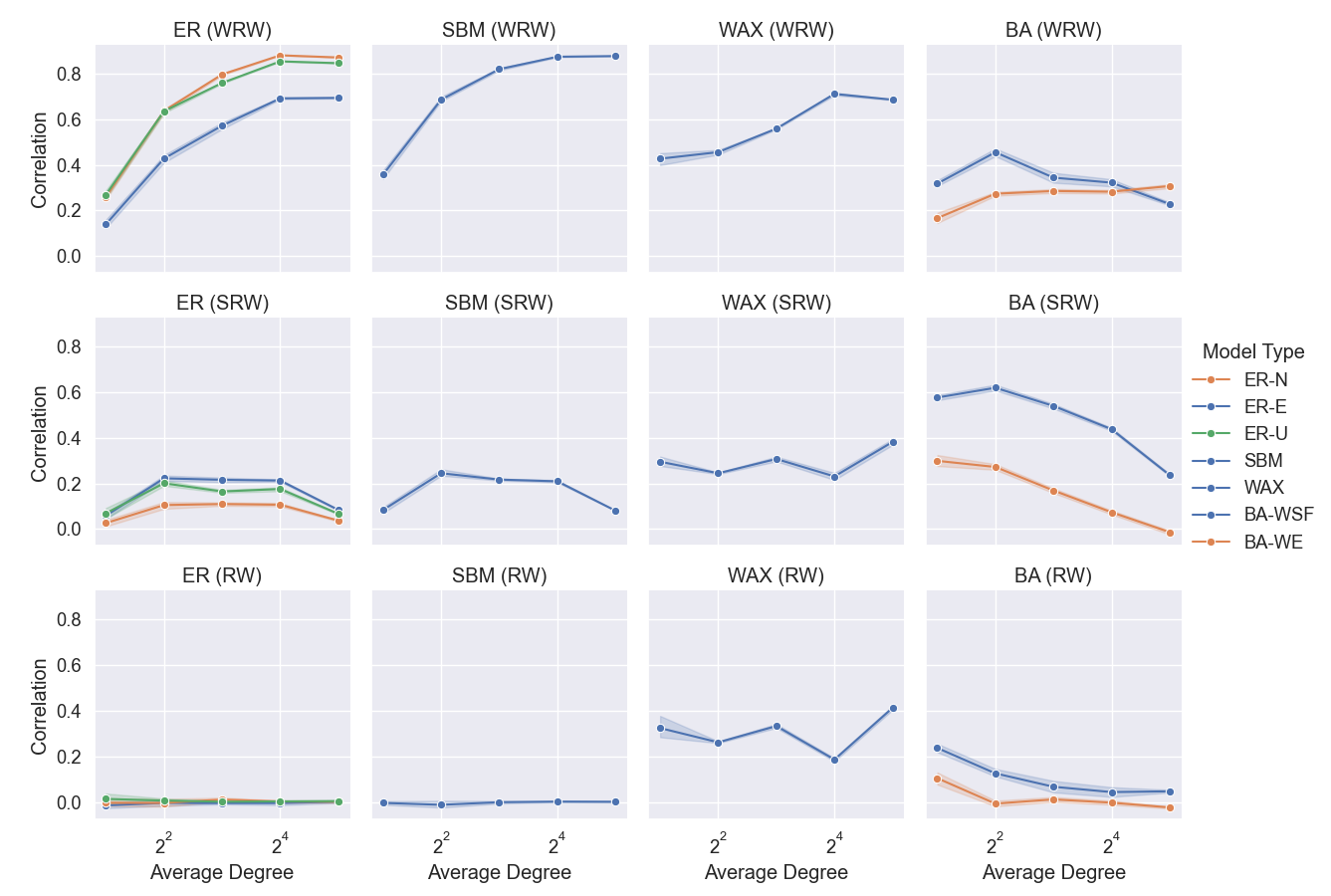}
\caption{Pearson correlation between cosine similarity of node embeddings and original edge weights across different graph models and random walk strategies. This analysis evaluates how well three types of random walks (WRW, SRW, RW) capture edge weight information in four graph models (ER, SBM, WAX, BA). Correlations were computed across multiple network instances, analyzing trends as the average node degree increased from 2 to 32 in powers of 2, with a fixed graph size of 2048 nodes. The experiment used a walk length of 128 and 16 walks per node.}
\label{fig:model_avg_degree}
\end{figure}

We now analyze the correlation's behavior as a function of the networks' average degree. This analysis was conducted on networks composed of 2048 nodes.
In Figure \ref{fig:model_avg_degree}, we have similar results as in Figure \ref{fig:model_graph_size}, with ER and SBM having correlations close to zero for the unweighted random walk (RW) but a small increase when using the node-strength (SRW) with values between 0.1 and 0.2, mostly represented by a plato. When using the weighted random walk (WRW), we can observe that correlations increase a lot along with the node degree, going from values below 0.4 to almost 0.9. For both parameters, graph size and node degree, as we increase their values we are also increasing the number of connections, increasing the weight redundancy and representing a wider variety of values and improving the efficiency of the walks in learning the distribution. Similar to the previous results, WAX had a high correlation for the unweighted and strength-based walks with values varying from 0.2 to 0.4, while an increase from 0.4 to 0.7 when using the weighted walk. WAX graphs are not very impacted by the presence of hubs as we will see for BA, making a noticiable improvement when adding the weight to the walk.

Finally, for the Barabási-Albert networks, we observe a downward trend for both the unweighted and strength-based random walks. This behavior is the inverse of what was observed when we fixed the average node degree and the correlation increased with the graph size. As the node degree increases, a few nodes are expected to accumulate a larger number of connections. Consequently, these nodes will have greater strengths, which can disrupt the transition probabilities and cause the random walk to concentrate around these hubs. This concentration leads to a poor representation of the entire network.
Similarly, the weighted random walk will be biased towards these hubs, having a bad representation of the weight distribution and producing bad inputs for the cosine similarity. Even for the unweighted random walk, we can see the same behavior of correlation dropping as the degree increases. In this scenario, we can see that initially, even without weights the cosine similarity can learn the weight representation, as the network structure is tightly linked to the weights. However, as the degree increases, a few nodes will have most of the connections, meaning that again the walk will concentrate in its surroundings. 

\subsubsection{Walks per Node and Walk Length}

\begin{figure}
\centering
\includegraphics[scale=0.5]{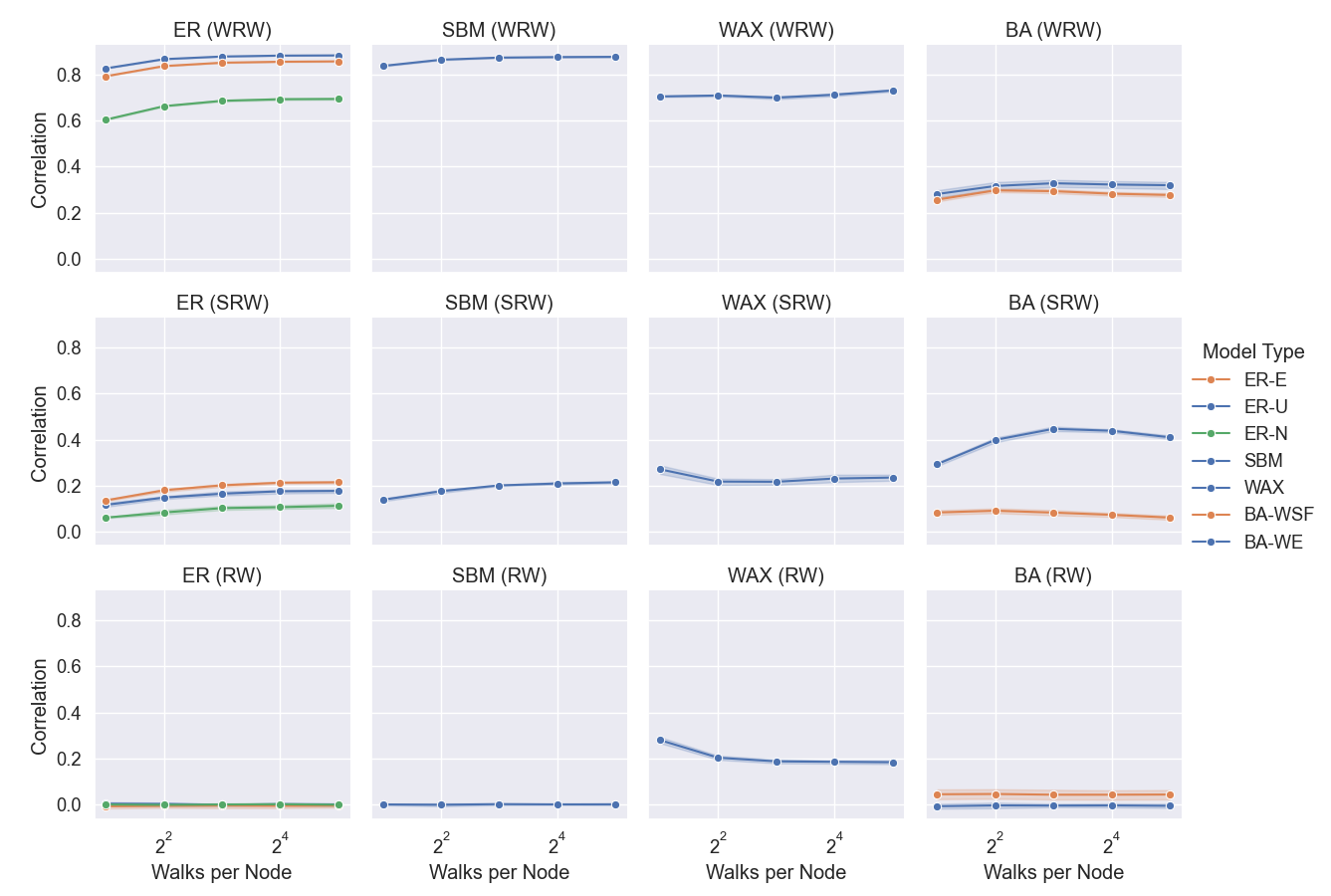}
\caption{Pearson correlation between cosine similarity of node embeddings and original edge weights across different graph models and random walk strategies. This analysis evaluates how well three types of random walks (WRW, SRW, RW) capture edge weight information in four graph models (ER, SBM, WAX, BA). Correlations were computed across multiple network instances, analyzing trends as the number of walks per node increased from 2 to 32 in powers of 2, with a fixed graph size of 2048 nodes, an average node degree of 16, and a walk length of 128.}
\label{fig:model_walks_per_node}
\end{figure}

\begin{figure}
\centering
\includegraphics[scale=0.5]{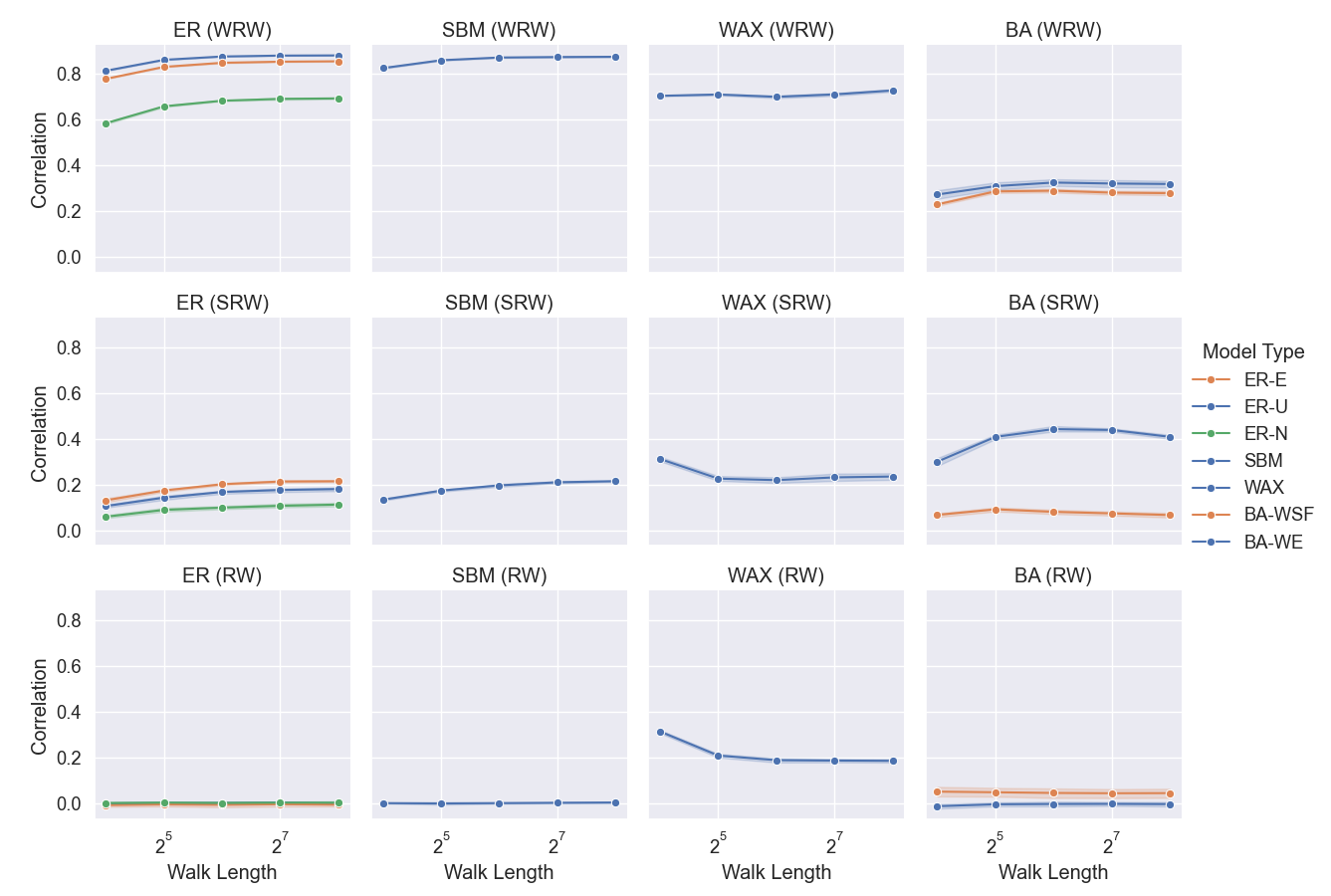}
\caption{Pearson correlation between cosine similarity of node embeddings and original edge weights across different graph models and random walk strategies. This analysis  evaluates how well three types of random walks (WRW, SRW, RW) capture edge weight information in four graph models (ER, SBM, WAX, BA). Correlations were computed across multiple network instances, analyzing trends as the walk length increased from 16 to 256, with a fixed graph size of 2048 nodes, an average node degree of 16, and 16 walks per node.}
\label{fig:model_walk_length}
\end{figure}

Figures \ref{fig:model_walks_per_node} and \ref{fig:model_walk_length} show the influence of the number of walks per node and the walk length on the observed correlation, respectively. For this analysis, we used a fixed graph size of 2048 nodes with an average node degree of 16. First, we kept the walk length constant at 128 while increasing the number of walks per node from 2 to 32. In the second case, we fixed the number of walks per node at 16 and varied the walk length from 16 to 256.

The results show a steady, albeit slight, improvement in performance for SRW and WRW, while RW remains unchanged. Both walks per node and walk length play a crucial role in capturing as many connections as possible from the network, ensuring that both local relationships between nodes and the global structure are learned during the walk process and propagated into the embeddings.
Increasing these parameters can enhance performance but also significantly raises computational cost, execution time, and memory usage. Therefore, to maintain efficiency while being resource-conscious, a few well-chosen iterations may be sufficient to extract meaningful information from the weight distribution. Ultimately, while walk properties have an influence, network properties have a much greater impact, where even a small change can lead to significant variations in correlation performance.

\subsection{Real Networks}

For the real-world networks, we selected 11 weighted graphs from the \emph{Netzschleuder network catalog}~\cite{tiago_p_peixoto_2020_7839981}. We
conducted the same analysis, as shown in Figure \ref{fig:methodology}. 
%
Additionally, we created a new version of each graph by randomly shuffling the weights while preserving the overall distribution and then applied the same random walks. 
This null-model approach enables us to isolate and quantify the contribution of the actual weight distribution compared to random noise. It also helps determine whether the structural topology alone (i.e., without meaningful weights) can drive strong correlations in embedding similarity. Recent work suggests that the performance of various biased random walks on downstream tasks often converges to similar outcomes due to shared structural information~\cite{vital2024comparing}, raising the question of whether walk bias and weight guidance truly contribute additional value or merely reinforce existing topological robustness. This experiment helps clarify whether observed correlations result from meaningful weight encoding or from structural redundancy.

The correlation results are presented in Table \ref{tab:realnetworks}, where the networks are sorted in descending order based on the correlation obtained using the Weighted Random Walk (WRW) on the original weights. For all experiments, we fixed the walk length to 128 and performed 16 walks per node.

The results reveal that not all graphs exhibited a strong correlation between cosine similarity and edge weights when using WRW on the original weight distribution. In particular, the \emph{New Zealand Collab}, \emph{Bible Nouns}, and \emph{Netscience} networks had correlations close to zero or even negative values when using unweighted (RW) or strength-based (SRW) walks. Another key observation is that correlations in real-world networks are generally lower than those observed in synthetic graph models. The highest correlation in this dataset was 0.446, observed in the \emph{SP high school diaries} network using WRW with the original weight distribution.

Interestingly, two graphs -- \emph{Sp High School Diaries} and \emph{Faculty Hiring US Academia} -- showed significant positive correlations even with the unweighted random walk (RW). Moreover, for both graphs, the correlation values remained similar when using shuffled weights, a pattern that mirrors what we previously observed in WAX model graphs. However, while the \emph{Sp High School Diaries} network showed a large improvement when switching from RW to WRW (0.227 with shuffled weights vs. 0.446 with original weights), the \emph{Faculty Hiring US Academia} network behaved differently. In this case, the highest correlation occurred with the strength-based walk (SRW) on shuffled weights (0.311), while WRW on original weights only reached 0.267.

\begin{table}[H]
    \centering
    \caption{Pearson correlation between cosine similarity and edge weights for 11 real-world graphs using different random walk strategies. The table compares the correlation values obtained from walks performed on networks with their original weight distribution and those where weights were shuffled while preserving the overall distribution. Three types of random walks were used: traditional random walk (RW), strength-based random walk (SRW), and weighted random walk (WRW). Higher correlation values in the original networks indicate that weight information is preserved during the walk, while lower values in the shuffled networks suggest a loss of meaningful weight-dependent structure.}
    \label{tab:realnetworks}
\begin{tabular}{l|cccccc|}
\cline{2-7}
\multicolumn{1}{c|}{\textbf{}}                                         & \multicolumn{6}{c|}{\textbf{Walk Type / Weights Type}}                                                                                                                                                                         \\ \cline{2-7} 
\multicolumn{1}{c|}{\textbf{}}                                         & \multicolumn{2}{c|}{\textbf{RW}}                                                & \multicolumn{2}{c|}{\textbf{SRW}}                                               & \multicolumn{2}{c|}{\textbf{WRW}}                          \\ \hline
\multicolumn{1}{|l|}{\textbf{Graph Name}}                              & \multicolumn{1}{c|}{\textbf{Shuffled}} & \multicolumn{1}{c|}{\textbf{Original}} & \multicolumn{1}{c|}{\textbf{Shuffled}} & \multicolumn{1}{c|}{\textbf{Original}} & \multicolumn{1}{c|}{\textbf{Shuffled}} & \textbf{Original} \\ \hline
\multicolumn{1}{|l|}{\emph{Sp High School Diaries}}               & \multicolumn{1}{c|}{0.240}             & \multicolumn{1}{c|}{0.224}             & \multicolumn{1}{c|}{0.218}             & \multicolumn{1}{c|}{0.279}             & \multicolumn{1}{c|}{0.227}             & 0.446             \\ \hline
\multicolumn{1}{|l|}{\emph{Celegansneural}}                          & \multicolumn{1}{c|}{0.060}             & \multicolumn{1}{c|}{0.051}             & \multicolumn{1}{c|}{0.144}             & \multicolumn{1}{c|}{0.221}             & \multicolumn{1}{c|}{0.029}             & 0.365             \\ \hline
\multicolumn{1}{|l|}{\emph{Fao Trade}}                              & \multicolumn{1}{c|}{0.028}             & \multicolumn{1}{c|}{0.042}             & \multicolumn{1}{c|}{0.022}             & \multicolumn{1}{c|}{-0.024}            & \multicolumn{1}{c|}{-0.002}            & 0.315             \\ \hline
\multicolumn{1}{|l|}{\emph{Celegans 2019 Hermaphrodite}} & \multicolumn{1}{c|}{0.049}             & \multicolumn{1}{c|}{0.050}             & \multicolumn{1}{c|}{0.044}             & \multicolumn{1}{c|}{0.092}             & \multicolumn{1}{c|}{0.038}             & 0.295             \\ \hline
\multicolumn{1}{|l|}{\emph{Celegans 2019 Male}}          & \multicolumn{1}{c|}{0.029}             & \multicolumn{1}{c|}{0.028}             & \multicolumn{1}{c|}{0.060}             & \multicolumn{1}{c|}{0.085}             & \multicolumn{1}{c|}{0.016}             & 0.284             \\ \hline
\multicolumn{1}{|l|}{\emph{Faculty Hiring US Academia}}           & \multicolumn{1}{c|}{0.205}             & \multicolumn{1}{c|}{0.202}             & \multicolumn{1}{c|}{0.311}             & \multicolumn{1}{c|}{0.245}             & \multicolumn{1}{c|}{0.166}             & 0.267             \\ \hline
\multicolumn{1}{|l|}{\emph{Cintestinalis}}                           & \multicolumn{1}{c|}{0.052}             & \multicolumn{1}{c|}{0.054}             & \multicolumn{1}{c|}{0.037}             & \multicolumn{1}{c|}{0.063}             & \multicolumn{1}{c|}{0.026}             & 0.258             \\ \hline
\multicolumn{1}{|l|}{\emph{Budapest Connectome All 200k}}         & \multicolumn{1}{c|}{0.065}             & \multicolumn{1}{c|}{0.064}             & \multicolumn{1}{c|}{0.055}             & \multicolumn{1}{c|}{0.067}             & \multicolumn{1}{c|}{0.027}             & 0.129             \\ \hline
\multicolumn{1}{|l|}{\emph{New Zealand Collab}}                    & \multicolumn{1}{c|}{-0.167}            & \multicolumn{1}{c|}{-0.166}            & \multicolumn{1}{c|}{-0.094}            & \multicolumn{1}{c|}{0.023}             & \multicolumn{1}{c|}{-0.150}            & 0.003             \\ \hline
\multicolumn{1}{|l|}{\emph{Bible Nouns}}                            & \multicolumn{1}{c|}{-0.159}            & \multicolumn{1}{c|}{-0.158}            & \multicolumn{1}{c|}{-0.080}            & \multicolumn{1}{c|}{-0.048}            & \multicolumn{1}{c|}{-0.157}            & -0.040            \\ \hline
\multicolumn{1}{|l|}{\emph{Netscience}}                              & \multicolumn{1}{c|}{-0.218}            & \multicolumn{1}{c|}{-0.225}            & \multicolumn{1}{c|}{-0.218}            & \multicolumn{1}{c|}{-0.234}            & \multicolumn{1}{c|}{-0.207}            & -0.084            \\ \hline
\end{tabular}
\end{table}

\subsection{Robustness and Threshold Analysis}

In this analysis, we investigate the impact of edge filtering by progressively removing low-weight edges from the network. This procedure simulates real-world scenarios involving data loss, noise, or uncertainty, where weak links are often pruned~\cite{amancio2015robustness}. By repeating the embedding process across a range of percentile-based threshold levels, we assess how the reduction in connectivity influences the ability of random-walk embeddings to preserve weight-related information. This analysis allows us to examine both the robustness and granularity of weight encoding -- highlighting whether specific thresholds represent critical tipping points at which the embedding effectiveness begins to deteriorate, or conversely, whether filtering enhances the signal by removing spurious connections~\cite{brito2020complex}.

Our approach is inspired by backbone extraction techniques in network science \cite{yassin2023evaluation}, where the goal is to simplify complex networks by pruning the weakest edges to reveal the underlying core structure. In this context, we aim to evaluate the robustness of weight-aware embeddings under edge reduction and determine how far weak edge removal can go before it begins to degrade the embedding capacity to capture meaningful weight patterns. Previous work has shown that emphasizing stronger connections can significantly improve community detection accuracy~\cite{amancio2015robustness,kovacs2024iterative}, suggesting that well-targeted edge filtering may sharpen the embedding representation of latent structural information. 

As described in the Section \ref{sec:complete}, we generated random, fully connected weighted graphs by initializing a matrix of random feature vectors, where each row represented a node and each column corresponded to a feature dimension. The weight of each edge was computed as the cosine similarity between the feature vectors of the corresponding node pair. 
This resulted in a complete weighted graph where all nodes were connected.
To reduce noise and study the effects of sparsification, we progressively removed the weakest connections. We did this by applying percentile-based thresholds, removing edges with weights below the 10th percentile all the way up to the 90th percentile of the distribution, in 10\% increments. 

\begin{figure}
\centering
\includegraphics[scale=0.48]{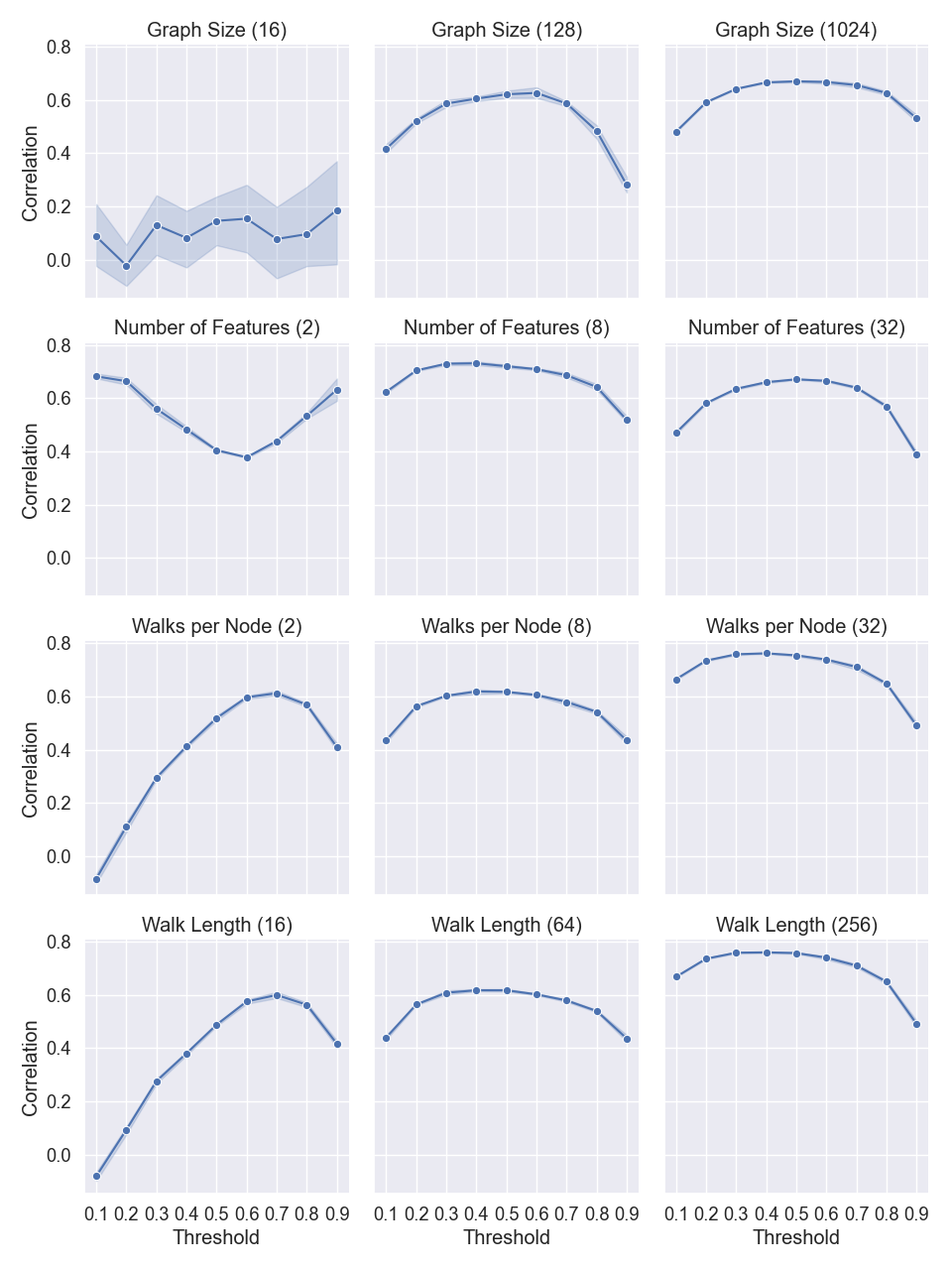}
\caption{Pearson correlation between cosine similarity of node embeddings and original edge weights for completed graphs, where edges with weights below a given percentile threshold were progressively removed. The x-axis represents the removal threshold (from 10\% to 90\%), and the y-axis shows the resulting correlation. Each subgraph varies a single graph or walk parameter while keeping the others fixed: (1st row) graph size, (2nd row) average node degree, (3rd row) number of walks per node, and (4th row) walk length. The analysis evaluates how correlation changes as low-weight edges are removed, revealing the sensitivity of different configurations to weight sparsification and the robustness of weight-aware embedding under varying network conditions.}
\label{fig:completednetwork}
\end{figure}

For each resulting pruned network, we applied a weighted random walk (WRW) to generate sequences of nodes, which were then used to train a skip-gram model and obtain node embeddings. These embeddings allowed us to evaluate the extent to which structural and weight information was preserved. As in previous experiments, we calculated the Pearson correlation between the cosine similarity of the embeddings and the original (pre-threshold) edge weights to assess embedding quality. To isolate the impact of key variables, we varied one parameter at a time -- such as graph size, number of features, number of walks per node, or walk length -- while keeping the others fixed. For visual clarity, we report only the results for the smallest, middle, and largest values of each parameter in Figure \ref{fig:completednetwork}, as the trends remained consistent across the full range.

The results reveal a consistent three-phase pattern in the behavior of correlation as the threshold increases. In the first phase, for thresholds between 0.1 and 0.3, we observe a steady rise in correlation. This suggests that removing the lowest-weight edges—often representing noise or weak relationships—helps denoise the graph. By retaining only stronger and more meaningful connections, the random walk more effectively captures the graph’s true structural and weight-based relationships, leading to higher-quality embeddings.
The second phase occurs between thresholds of 0.3 and 0.7, where the correlation stabilizes into a plateau. At this stage, further edge removal does not yield additional performance benefits and, in some configurations, may cause a slight decline. This suggests that the graph reaches a critical sparsity level where remaining edges are already highly informative, and additional pruning may begin to affect its structural integrity. Finally, in the third phase --beyond the 0.7 threshold -- we observe a sharp drop in correlation. As the network becomes overly sparse, important structural information is lost, and the resulting walks are no longer sufficient to capture the underlying weight distribution. In many cases, performance at this stage drops below that of the original, unpruned graph, confirming that excessive pruning significantly degrades embedding quality

Each parameter configuration exhibits this general pattern, but with subtle differences in the early phase. For example, when using a small number of walks per node or a short walk length, the initial correlation is very low due to insufficient sampling of the graph. As these parameters increase, the walks cover more of the network, leading to higher correlation scores—until the point of over-pruning, where performance declines again. The number of features used to generate the initial cosine similarity weights introduces a unique behavior, often forming a U-shaped curve. With very few features, embeddings lack expressiveness; increasing the number initially reduces correlation as noise is still present, but further increases allow for clearer separation between strong and weak connections, resulting in improved correlation at higher thresholds. In contrast, the variation in graph size reveals a different trend. Particularly for smaller graphs, correlation remains low and unstable across all thresholds, likely due to limited node diversity and fewer opportunities for the walk to explore meaningful patterns. This behavior was not observed with the other parameters, which showed clearer improvements in performance before declining at high thresholds.

Ultimately, these findings emphasize the delicate balance between removing noise and preserving essential structure in weighted graphs. Across all parameter settings, the ability to retrieve meaningful weight information from node embeddings is tightly coupled with the underlying structure of the graph. Initially, low-weight edges act as noise, obscuring the walk’s ability to encode relevant relationships. Removing these edges improves performance up to an optimal point. Beyond that, however, over-pruning reduces the connectivity of the graph and impairs the quality of the random walks, leading to degraded embeddings. This behavior highlights the importance of tuning the edge removal threshold to strike a balance between denoising and structural preservation in graph representation learning.

\section{Conclusion}
\label{chapter:conclusion}

This study assessed how various random walk strategies incorporate edge weight information within graph models and real-world networks. The results provide critical insights into the influence of graph structure, weight distribution, and walk parameters on the generation of meaningful node embeddings. The methodology involved correlating the original edge weight information with the similarity of nodes as measured by their random walk-based embeddings.

For graph models, the Weighted Random Walk consistently achieved the highest correlations,
particularly in ER, SBM, and WAX models, with values exceeding 0.9 in some cases. However, in BA models, the increasing presence of hubs distorted the weight distribution, leading to lower correlations. RW, as expected, demonstrated poor correlation across all models, except for Waxman (WAX), where the network’s spatial-based weight assignment inherently preserved weight relationships, yielding a correlation around 0.40. SRW improved correlation compared to RW, particularly in the Weighted Scale-Free Barabási-Albert (BA-WSF) model, where correlation steadily increased as graph size grew, reaching 0.5 at 4096 nodes.

Graph size and node degree played a crucial role in performance. Increasing graph size improved correlation in WRW for most models, but in BA models, large networks introduced hubs that biased walk sampling, reducing effectiveness. Similarly, increasing node degree improved WRW correlation up to 0.9, as it enhanced weight redundancy and connectivity. However, in BA models, higher node degrees concentrated walks around hubs, leading to lower correlations for all methods. Walk parameters, such as walk length and the number of walks per node, had a marginal but consistent effect, with longer and more frequent walks slightly improving SRW and WRW, while RW remained unchanged. Although extensive walks enhanced learning, they also increased computational costs, requiring a balance between accuracy and efficiency.

When applied to real-world networks, WRW again outperformed the other walk strategies, but overall, correlations were lower than in synthetic models. The highest correlation (0.446) was observed in the \emph{SP High School Diaries} network using WRW on original weights. However, some networks, such as \emph{New Zealand Collab}, \emph{Bible Nouns}, and \emph{Netscience}, exhibited near-zero or even negative correlations, indicating that their weight structures do not align well with node embeddings. Notably, the \emph{Faculty Hiring US Academia} network achieved its best performance with SRW on shuffled weights (0.311) rather than WRW on original weights. This suggests that some networks benefit from weight randomization by preventing walk concentration on high-strength nodes, a phenomenon also observed in BA-WSF models.

Our analysis of completed graphs revealed that removing low-weight edges can significantly improve node embedding quality. This process reduces noise and strengthens the correlation between cosine similarity and original edge weights.
However, this improvement holds only up to an optimal threshold -- typically between the 30th and 60th percentiles -- beyond which over-pruning degrades the network structure and harms performance. This highlights the importance of balancing noise reduction with structural integrity to effectively preserve weight information in graph embeddings.

Our findings reinforce that weight-aware random walks (WRW) are essential for effectively preserving edge weight relationships in node embeddings. However, graph structure significantly influences performance. For instance, in completed graphs and with models like BA-WSF and WAX, weight relationships are naturally embedded more effectively than with ER and SBM models. High node degrees and the presence of hubs can also distort weight representation, particularly in scale-free networks.
Real-world networks exhibit more variability, with some benefiting from weight shuffling, which suggests that optimal performance may require network-dependent strategies. These insights are particularly relevant for applications in transportation modeling, recommendation systems, and financial networks, where both graph structure and edge weights convey meaningful information. Future research could investigate hybrid approaches that adapt walk strategies based on network topology to further improve embedding quality.

\section*{Acknowledgments}

This study was supported by the National Council for Scientific and Technological Development – CNPq (Grant number 304189/2025-1).
Filipi N. Silva thanks AFOSR \#FA9550-19-1-0391 for the financial support.

\bibliographystyle{apalike}

\bibliographystyle{abbrvnat}

\newpage

\end{document}